\documentclass[12pt,a4paper,reqno]{amsart}
\usepackage[all]{xy}
\usepackage{amssymb}
\usepackage[latin1]{inputenc}
\usepackage{color}
\usepackage[margin=2.5cm,papersize={19.75cm,28cm}]{geometry}
\usepackage{hyperref}
\usepackage{eucal}
\usepackage{epsfig}
\usepackage{graphicx}
\usepackage{bm}

\newtheorem{definition}{Definition}[section]

\newtheorem{theorem}[definition]{Theorem}

\newtheorem{remarkth}[definition]{Remark}

\renewcommand{\emph}[1]{{\bfseries\itshape{#1}}}

\newcommand{\R}{\mathbb{R}}      %Numeros reales
\newcommand{\N}{\mathbb{N}}      %Numeros naturales
\begin{document}
\title[Higher-order variational problems on Lie groups]{On the geometry of higher-order variational problems on Lie groups}

\author[L.\ Colombo]{Leonardo\ Colombo}
\address{Leonardo Colombo:
Instituto de Ciencias Matem\'aticas (CSIC-UAM-UC3M-UCM),
Campus de Cantoblanco, UAM
C/ Nicolas Cabrera, 15 - 28049 Madrid (SPAIN),
28006 Madrid, Spain}
\email{leo.colombo@icmat.es}

\author[D.\ Mart\'\i n de Diego]{David Mart\'\i n de Diego}
\address{David Mart\'\i n de Diego:
Instituto de Ciencias Matem\'aticas (CSIC-UAM-UC3M-UCM),
Campus de Cantoblanco, UAM
C/ Nicolas Cabrera, 15 - 28049 Madrid (SPAIN),
28006 Madrid, Spain}
\email{david.martin@icmat.es}

\thanks{\noindent This work has been partially supported by MEC (Spain)
Grants   MTM 2010-21186-C02-01, MTM2009-08166-E,   project ``Ingenio
Mathematica" (i-MATH) No. CSD 2006-00032 (Consolider-Ingenio 2010), and IRSES-project ``Geomech-246981''. L.Colombo also wants to thank CSIC for a JAE-Pre grant}


\begin{abstract}
In this paper, we describe a geometric setting for higher-order lagrangian problems on Lie groups. Using left-tri\-via\-li\-za\-tion of the higher-order tangent bundle of a Lie group and an adaptation of the classical Skinner-Rusk formalism,  we deduce an intrinsic framework for this type of dynamical systems. Interesting applications as, for instance, a geometric derivation  of the higher-order Euler-Poincar\'e equations, optimal control of underactuated control systems whose configuration space is a Lie group are shown, among others, along the paper.
\end{abstract}

\maketitle

\section{Introduction}
In 1901 \cite{Po1901},  H. Poincar\'e variationally deduced   the equations of motion of a mechanical system specified by a lagrangian $l: {\mathfrak g}\to \R$ where ${\mathfrak g}$ is the Lie algebra of a Lie group $G$. These equations are actually known as the Euler-Poincar\'e equations. There mainly appear as the reduction of a lagrangian $L: TG\to \R$ being  left or right-invariant. This procedure is called Euler-Poincar\'e reduction (see \cite{holm,tudor}).
The Euler-Poincar\'e  equations are
\begin{equation}\label{Euler-Poincare}
\frac{d}{dt}\frac{\delta  l}{\delta \xi}-ad_{\xi}^*\frac{\delta l}{\delta \xi}=0
\end{equation}
where $ad_\xi^*$ is the dual operator of the adjoint endomorphism $ad_{\xi}: {\mathfrak g}\to {\mathfrak g}$ defined by $ad_{\xi} \tilde{\xi}=[\xi, \tilde{\xi}]$ where $\xi, \tilde{\xi}\in {\mathfrak g}$. Typically, it is used  the functional derivative notation $ \delta l/\delta \xi$ since the equations are also valid for infinite dimensional Lie algebras. As particular examples, these equations include the equations for rigid bodies and fluids, but in the latter
case, one must use infinite dimensional Lie algebras.
In the variational deduction of the Equations (\ref{Euler-Poincare}), it is necessary a careful analysis of the admissible infinitesimal variations to deduce the equations. More precisely, the variations are obtained by a reduction procedure of the admissible variations of the unreduced lagrangian $L: TG\to \R$.
Of course, using the corresponding Legendre transformation, it is possible to rewrite Equations (\ref{Euler-Poincare}) as the Lie-Poisson equations on  ${\mathfrak g}^*$.
Observe that, as an essential feature, Equations (\ref{Euler-Poincare}) involve half of the degrees of freedom as compared with the usual Euler-Lagrange equations for $L: TG\to \R$ and, moreover, Equations (\ref{Euler-Poincare}) are first-order differential equations while the standard Euler-Lagrange equations are second-order ones.

Very recently, and from different motivations, there appears a considerable interest on the extension of Equation (\ref{Euler-Poincare}) to higher-order mechanics (see, for instance, \cite{CSC},\cite{LR1} as typical references for higher-order mechanics on tangent bundles). Our main objective in the paper is to characterize geometrically the equations of motion for an optimal control problem of a possibly underactuated  mechanical system. In this last system, the trajectories are ``parameterized" by the admissible controls and the necessary conditions for extremals in the optimal control problem are expressed using a ``pseudo-hamiltonian formulation" based on the Pontryaguin maximun principle or an appropriate variational setting using some smoothness conditions \cite{acccms}. Many of the concrete examples under study have additional geometric properties, as for instance, that the configuration space is not only a differentiable manifold but  it also  has a compatible structure of group, that is, the configuration space is a Lie group. In this paper, we will take advantage of this property   to give a closed and intrinsic form of the equations of motion of our initial optimal control problem. For it, we will use extensively the Skinner-Rusk formalism which combines simultaneously some features of the lagrangian and hamiltonian classical formalisms and, as we will show, it is the adequate space to study the problems that we want to characterize \cite{Skinner-Rusk}.
Other interesting characteristic of the problems under study is that, for their characterization, it is necessary to use higher-order mechanics, that is, the phase space on the lagrangian side has coordinates which specify positions, velocities and accelerations. In the applications, we will use only second-order tangent bundles, but since the extension to $k$-order tangent bundles does not offer special difficulties we will develop our geometric theory  in this last case.

Moreover, in a recent paper \cite{GHMRV10}, the authors study from a pure variational point of view,  invariant higher-order variational problems with the idea to analyze  higher-order geometric $k$-splines with the application to image analysis. The obtain as main result a version of the higher-order
Euler-Poincar\'e equations.
We will show in our paper that our analysis is complementary to the one in \cite{GHMRV10}. Assuming that our initial problem is invariant we deduce geometrically the same equations than the authors, but they use mainly variational techniques.

The paper is structured as follows. In Section \ref{section1}, we introduce some geometric constructions which are used along the paper. In particular, the Euler-Arnold equations for a hamiltonian system defined on the cotangent bundle of a Lie group and their extension to higher-order cases.
In Section  \ref{subsection2}, we define the Pontryaguin bundle  $G\times k{\mathfrak g}\times k{\mathfrak g}^*$ where we introduce the dynamics using a presymplectic hamiltonian formalism. We deduce the $k$-order Euler-Lagrange equations and, as a particular example, the $k$-order Euler-Poincar\'e equations. Since the dynamics is presymplectic it is necessary to analyze the consistency of the dynamics using a constraint algorithm \cite{GoNe,GoNesHinds78}. Section \ref{subsection3} is devoted to the case of constrained dynamics. We show that our techniques are easily adapted to this particular case. As an illustration of the applicability of our setting, we analyze the case of underactuated control of mechanical systems in Section \ref{section4} and, as a particular example, a family of underactuated problems for the rigid body on $SO(3)$.

\section{Geometric preliminaries}\label{section1}

\subsection{Euler-Arnold equations} (See, for instance, \cite{Chusman-Bates})
Let $G$ be a Lie group. Consider the left-multiplication on itself
\[
G\times G\longrightarrow G\; ,\qquad (g, h)\rightarrow \pounds_g(h)=gh\; .
\]
Obviously $\pounds_g$ is a diffeomorphism. (The same is valid for the right-translation, but in the sequel we only work with the left-translation, for sake of simplicity).

This left multiplication allows us to trivialize the tangent bundle $TG$ and the cotangent bundle $T^*G$ as follows
\begin{eqnarray*}
TG&\to&G\times {\mathfrak g}\, ,\qquad (g, \dot{g})\longmapsto (g, g^{-1}\dot{g})=(g, T_g\pounds_{g^{-1}}\dot g)=(g, \xi)\; ,\\
T^*G&\to&G\times {\mathfrak g}^*,\qquad  (g, \alpha_g)\longmapsto (g, T^*_e\pounds_g(\alpha_g))=(g, \alpha)\; ,
\end{eqnarray*}
 where ${\mathfrak g}=T_eG$ is the Lie algebra of $G$ and $e$ is the neutral element of $G$.
In the same way, we have the following identifications:  $TTG\equiv G\times 3{\mathfrak g}$, $T^*TG=G\times {\mathfrak g}\times 2{\mathfrak g}^*$.
$TT^*G=G\times {\mathfrak g}^*\times {\mathfrak g}\times {\mathfrak g}^*$ and $T^*T^*G=G\times 3{\mathfrak g}^*$.

Using this left trivialization it is possible to write the classical hamiltonian equations for a hamiltonian function $H: T^*G\rightarrow \R$
from a different and interesting perspective.

For instance, it is easy to show that the canonical structures of the cotangent bundle: the Liouville 1-form $\theta_G$ and the canonical symplectic 2-form $\omega_G$, are now rewritten using this left-trivialization as follows:
\begin{eqnarray}
(\theta_G)_{(g, \alpha)}(\xi_1, \nu_1)&=&\langle \alpha, \xi_1\rangle\; ,\\
(\omega_G)_{(g, \alpha)}\left( (\xi_1, \nu_1), (\xi_2, \nu_2)\right)&=&-\langle \nu_1, \xi_2\rangle + \langle \nu_2, \xi_1\rangle+\langle\alpha, [\xi_1, \xi_2]\rangle\; ,\label{omega}
\end{eqnarray}
with $(g, \alpha)\in G\times {\mathfrak g}^*$,
where $\xi_i\in {\mathfrak g}$ and $\nu_i\in {\mathfrak g}^*$, $i=1, 2$  and we have used the previous identifications. Observe that we are identifying the elements of $T_{\alpha_g}T^*G$ with the pairs $(\xi, \nu)\in {\mathfrak g}\times {\mathfrak g}^*$.

Therefore given the hamiltonian $H: T^*G\equiv G\times {\mathfrak g}^*\longrightarrow \R$, we compute
\begin{equation}
dH_{(g, \alpha)}(\xi_2, \nu_2)=\langle  \pounds_g^*\left(\frac{\delta H}{\delta g}(g, \alpha)\right), \xi_2\rangle+ \langle \nu_2, \frac{\delta H}{\delta \alpha}(g, \alpha)\rangle\; , \label{hami-0}
\end{equation}
since $\frac{\delta H}{\delta \alpha}(g, \alpha)\in {\mathfrak g}^{**}={\mathfrak g}$.

We now derive the Hamilton's equations which are satisfied by the integral curves of the Hamiltonian vector field $X_H$ on $T^*G$. After left-trivialization,  $X_H(g, \alpha)=(\xi_1, \nu_1)$ where $\xi_1\in {\mathfrak g}$ and $\nu_1\in {\mathfrak g}^*$ are elements to be determined using the Hamilton's equations
\[
i_{X_H}\omega_G=dH\; .
\]
 Therefore, from expressions (\ref{omega}) and (\ref{hami-0}) we deduce that
\begin{eqnarray*}
\xi_1&=&\frac{\delta H}{\delta \alpha}(g, \alpha)\; ,\\
\nu_1&=&-\pounds_g^*\left(\frac{\delta H}{\delta g}(g, \alpha)\right)+ad_{\xi_1}^*\alpha\; .
\end{eqnarray*}
In other words, taking $\dot{g}=g\xi_1$ we obtain the \emph{Euler-Arnold equations}:
\begin{eqnarray*}
\dot{g}&=&T_e\pounds_g(\frac{\delta H}{\delta \alpha}(g, \alpha))\equiv g \frac{\delta H}{\delta \alpha}(g, \alpha)\; ,\\
\dot{\alpha}&=&-\pounds_g^*\left(\frac{\delta H}{\delta g}(g, \alpha)\right)+ad_{\frac{\delta H}{\delta \alpha}(g, \alpha)}^*\alpha\; .
 \end{eqnarray*}
If the Hamiltonian  is left-invariant, that, is $h: {\mathfrak g}^*\to \R$ where $h(\alpha)=H(e, \alpha)$ then we deduce that:
\begin{eqnarray*}
\dot{g}&=& g \frac{\delta h}{\delta \alpha}\, ,\\
\dot{\alpha}&=&ad_{{\delta h}/{\delta \alpha}}^*\alpha\; .
\end{eqnarray*}
The last equation is known as the \emph{Lie-Poisson equations} for a hamiltonian $h: {\mathfrak g}^*\to \R$.

\subsection{Higher-order tangent bundles}

In this section we recall some basic facts of the higher-order
tangent bundle theory. Along the section  we will particularize this construction to the case when the configuration space is a Lie group $G$.   For more details see \cite{CSC,LR1}.
%%%%%%%%%%%%%%%%%

Let $Q$ be a  differentiable manifold of dimension $n$. It is possible to introduce an equivalence relation
 in the set $C^{k}(\R, Q)$ of $k$-differentiable
curves from $\R$ to $Q$. By definition,  two given curves in $Q$,
$\gamma_1(t)$ and $\gamma_2(t)$,
where $t\in (-a, a)$ with $a\in \R$
have contact of order  $k$ at $q_0 = \gamma_1(0) = \gamma_2(0)$ if
there is a local chart $(\varphi, U)$ of $Q$ such that $q_0 \in U$
and
$$\frac{d^s}{dt^s}\left(\varphi \circ \gamma_1(t)\right){\big{|}}_{t=0} =
\frac{d^s}{dt^s} \left(\varphi
\circ\gamma_2(t)\right){\Big{|}}_{t=0}\; ,$$ for all $s = 0,...,k.$ This
is a well defined equivalence relation
in $C^{k}(\R,Q)$
and the equivalence class of a  curve $\gamma$ will be denoted by
$[\gamma ]_0^{(k)}.$ The set of equivalence classes will be
denoted by $T^{(k)}Q$
and it is not hard to show that it has a natural structure of differentiable manifold. Moreover, $ \tau_Q^k  : T^{(k)} Q
\rightarrow Q$ where $\tau_Q^k \left([\gamma]_0^{(k)}\right) =
\gamma(0)$ is a fiber bundle called the \emph{tangent bundle of
order $k$} of $Q.$

       In the case when the manifold  $Q$ has a Lie group structure, we will denote $Q=G$ and  we can also use  the left trivialization to identify the higher-order tangent bundle $T^{(k)} G$ with $G\times k{\mathfrak g}$. That is, if $g: I\rightarrow G$ is a curve in $C^{(k)}(\R, G)$:
\[
\begin{array}{rrcl}
\Upsilon^{(k)}:& T^{(k)}G&\longrightarrow& G\times k{\mathfrak g}\\
       & [g]_0^{(k)}&\longmapsto&(g(0), g^{-1}(0)\dot{g}(0), \frac{d}{dt}\Big|_{t=0}(g^{-1}(t)\dot{g}(t)), \ldots, \frac{d^{k-1}}{dt^{k-1}}\Big|_{t=0}(g^{-1}(t)\dot{g}(t)))
       \end{array}
\]
It is clear that $\Upsilon^{(k)}$ is a diffeomorphism

We will denote by $\xi(t)=g^{-1}(t)\dot{g}(t)$. Therefore
\[
\Upsilon^{(k)}([g]_0^{(k)})=(g, \xi, \dot{\xi}, \ldots, \xi^{(k-1)})\; ,
\]
where
\[
\xi^{(l)}(t)=\frac{d^{l}}{dt^{l}}(g^{-1}(t)\dot{g}(t)), \qquad 0\leq l\leq k-1
\]
and $g(0)=g, \xi^{(l)}(0)=\xi^{(l)}, 0\leq l\leq k-1$.
We will indistinctly use the notation $\xi^{(0)}=\xi$, $\xi^{(1)}=\dot{\xi}$, where there is not danger of confusion.

We   may also
define the surjective mappings $\tau_G^{(l,k)} : T^{(k)} G
\rightarrow T^{(l)} G,$ for $l\leq k$, given by
$\tau_G^{(l,k)}\left([g]_0^{(k)}\right) = [g]_0^{(l)}.$
With the previous identifications we have that
\[
\tau_G^{(l,k)} (g(0), \xi(0), \dot{\xi}(0), \ldots, \xi^{(k-1)}(0))=(g(0), \xi(0), \dot{\xi}(0), \ldots, \xi^{(l-1)}(0))
\]
It is easy to see that $T^{(1)} G \equiv G\times {\mathfrak g}$,  $T^{(0)} G \equiv G$ and $\tau_G^{(0,k)}=\tau_G^k$.

%Given a differentiable function $f: G\longrightarrow \R$ and $l
%\in \{0,...,k\}$, its $l$-lift $f^{(l, k)}$ to $T^{(k)}G$, $0\leq
%l\leq k$, is the
%%
%differentiable
%%
%function defined as
%\[
%f^{(l, k)}([g]^{(k)}_0)=\frac{d^l}{dt^l}{\Big{|}}_{t=0}
%\left(f \circ g(t)\right)\; .
%\]
%
%
%{} From a local chart $(q^i)$ on a neighborhood $U$ of $Q$, it is possible to induce local coordinates
% $(q^{(0)i},q^{(1)i},\dots,q^{(k)i})$ on
%$T^{(k)}U=(\tau_Q^k)^{-1}(U)$, where $q^{(s)i}=(q^i)^{(s,k)}$ if
%$0\leq s\leq k$. Sometimes, we will use the standard  conventions, $q^{(0)i}\equiv q^i$, $q^{(1)i}\equiv \dot{q}^i$ and
%$q^{(2)i}\equiv \ddot{q}^i$.
%
%
%Given a vector field $X$ on $Q$, we define its $k$-lift $X^{(k)}$
%to $T^{(k)}Q$ as the unique vector field on $T^{(k)}Q$ satisfying the following
%identities
%\[
%X^{(k)}(f^{(l, k)})=(X(f))^{(l,
%k)}\, , \hspace{1.2cm} 0\leq l\leq k\, ,
%\]
%for all differentiable function $f$ on $Q$.
% In coordinates, the $k$-lift of a
%vector field $\displaystyle{X=X^i\frac{\partial}{\partial q^i}}$
%is
%\[
%X^{(k)}=(X^i)^{(s, k)}\frac{\partial}{\partial
%q^{(s)i}}\; .
%\]

%
Now, we consider the canonical immersion
$j_k: T^{(k)}G\rightarrow T(T^{(k-1)} G)$ defined as
$j_k([g]_0^{(k)})=[{g}^{(k-1)}]_0^{(1)}$, where
${g}^{(k-1)}$ is the lift of the curve
$g$ to $T^{(k-1)}G$; that is,  the curve ${g}^{(k-1)}: \R\rightarrow
T^{(k-1)}G$ is given by $g^{(k-1)}(t)=[g_t]_0^{(k-1)}$
where $g_t(s)=g(t+s)$. Using the identification given by $\Upsilon^{(k)}$ we have that:
\[
\begin{array}{rrcl}
j^{(k)}:& G\times k{\mathfrak g}&\longrightarrow& G\times (2k-1){\mathfrak g}\\
       &(g, \xi, \dot{\xi}, \ldots, \xi^{(k-1)})&\longmapsto&(g, \xi, \dot{\xi},  \ldots, \xi^{(k-2)}; \; \xi,  \dot{\xi}, \ldots, \xi^{(k)})
       \end{array}
\]
where we identify $T(T^{(k-1)} G)\equiv T(G\times (k-1){\mathfrak g})\equiv G\times (2k-1){\mathfrak g}$, in the natural way.

\subsection{Higher-order Euler-Arnold equations  on $T^*(T^{(k-1)}G)$}

Combining the results of the two previous subsections we have that
\[
T^*(T^{(k-1)}G)\equiv T^*(G\times (k-1){\mathfrak g})\equiv T^*G\times (k-1)T^*{\mathfrak g}\equiv G\times (k-1){\mathfrak g}\times k {\mathfrak g}^*\; .
\]

For developing our geometric formalism for higher-order variational problems on Lie groups we need to equip the previous space   with a symplectic structure. Thus, we construct a  Liouville 1-form $\theta_{G\times (k-1){\mathfrak g}}$ and a canonical symplectic 2-form $\omega_{G\times (k-1){\mathfrak g}}$ after the left-trivialization  that we are using.
Denote by ${\bm \xi}\in (k-1){\mathfrak g}$ and ${\bm \alpha}\in k{\mathfrak g}^*$ with components
${\bm \xi}=(\xi^{(0)}, \ldots, \xi^{(k-2)})$ and ${\bm \alpha}=(\alpha_0,\ldots,\alpha_{k-1})$.
Then, after a straightforward computation we deduce that
\begin{eqnarray*}
(\theta_{G\times (k-1){\mathfrak g}})_{(g,{\bm \xi},  {\bm   \alpha})}({\bm \xi}_1, {\bm \nu}^1)&=&\langle {\bm \alpha}, {\bm \xi}_1\rangle\, ,\\
(\omega_{G\times (k-1){\mathfrak g}})_{(g,{\bm \xi},  {\bm   \alpha})}\left( ({\bm \xi}_1, {\bm \nu}^1), ({\bm \xi}_2, {\bm \nu}^2)\right)&=&-\langle {\bm \nu}^1, {\bm \xi}_2\rangle + \langle {\bm \nu}^2, {\bm \xi}_1\rangle+ \langle\alpha_0, [ \xi^{(0)}_1, \xi^{(0)}_2]\rangle\label{omega-1}\\
\hspace{-1cm}&=&-\sum_{i=0}^{k-1}\left[\langle {\nu}^1_{(i)}, {\xi}^{(i)}_2\rangle + \langle {\nu}^2_{(i)}, {\xi}^{(i)}_1\rangle\right]+ \langle\alpha_0, [\xi^{(0)}_1, \xi^{(0)}_2]\rangle\nonumber
\end{eqnarray*}
where ${\bm\xi}_a\in k{\mathfrak g}$ and ${\bm \nu}^a\in k{\mathfrak g}^*$, $a=1, 2$ with components ${\bm\xi}_a=(\xi^{(i)}_a)_{0\leq i\leq k-1}$ and
${\bm\nu}^a=(\nu_{(i)}^a)_{0\leq i\leq k-1}$ where each component $\xi^{(i)}_a\in {\mathfrak g}$ and $\nu_{(i)}^a\in {\mathfrak g}^*$. Observe that $\alpha_0$ comes from the identification $T^*G=G\times {\mathfrak g}^*$.

Given  the hamiltonian $H: T^*T^{(k-1)}G\equiv G\times (k-1){\mathfrak g}\times k {\mathfrak g}^*\longrightarrow \R$, we compute
\begin{eqnarray*}
dH_{(g,{\bm \xi},  {\bm   \alpha})}({\bm \xi}_2, {\bm \nu}^2)&=&\langle \pounds_g^*\left(\frac{\delta H}{\delta g}{(g,{\bm \xi},  {\bm   \alpha})}\right), {\xi}^{(0)}_2\rangle +\sum_{i=0}^{k-2}\langle \frac{\delta H}{\delta \xi^{(i)}}{(g,{\bm \xi},  {\bm   \alpha})}, {\xi}^{(i+1)}_2\rangle\\
&&+
 \langle {\bm \nu}^2, \frac{\delta H}{\delta {\bm \alpha}}{(g,{\bm \xi},  {\bm   \alpha})}\rangle \label{hami-1}
\end{eqnarray*}

As in the first subsection, we can  derive the Hamilton's equations which are satisfied by the integral curves of the Hamiltonian vector field $X_H$  defined by $X_H(g,{\bm \xi},  {\bm   \alpha})=({\bm \xi}_1, {\bm\nu}^1)$. Therefore,  we deduce that
\begin{eqnarray*}
{\bm \xi}_1&=&\frac{\delta H}{\delta\bm \alpha}(g,{\bm \xi},  {\bm   \alpha})\; ,\\
\nu_{(0)}^1&=&-\pounds_g^*\left(\frac{\delta H}{\delta g}(g,{\bm \xi},  {\bm   \alpha})\right)+ad_{\xi^{(0)}_1}^*\alpha_{0}\; ,\\
\nu_{(i+1)}^1&=&-\frac{\delta H}{\delta \xi^{(i)}}(g,{\bm \xi},  {\bm   \alpha}),\qquad 0\leq i\leq k-2\; .
\end{eqnarray*}
In other words, taking $\dot{g}=g\xi^{(0)}$ we obtain the \emph{higher-order Euler-Arnold equations}:
\begin{eqnarray*}
\dot{g}&=&g\frac{\delta H}{\delta \alpha_{0}}(g,{\bm \xi},  {\bm   \alpha})\; ,\\
\frac{d\xi^{(i)} }{dt}&=&\frac{\delta H}{\delta \alpha_{i}}(g,{\bm \xi},  {\bm   \alpha}), \qquad 1\leq i\leq k-1\; ,\\
\frac{d{\alpha}_{0}}{dt}&=&-\pounds_g^*\left(\frac{\delta H}{\delta g}(g,{\bm \xi},  {\bm   \alpha})\right)+ad_{{\delta H}/{\delta \alpha_{0}}}^*\alpha_{0}\, ,\\
\frac{d{\alpha}_{i+1}}{dt}&=&-\frac{\delta H}{\delta \xi^{(i)}}(g,{\bm \xi},  {\bm   \alpha}),\qquad 0\leq i\leq k-2\; .
 \end{eqnarray*}

%We use the map $j_{k}$ to construct the differential operator
%$d_{T}$ which maps a function $f$ on $T^{(k)}Q$ into a function
%$d_{T}f$ on $T^{(k+1)}Q$
%
%$$d_{T}f([\gamma]_{0}^{k+1}) =
%j_{k+1}([\gamma]_{0}^{k+1})(f)\; .$$

\section{On the geometry of higher-order variational problems on Lie groups}\label{section2}

In this section, we  describe the main results of the paper. First, we intrinsically  derive  the equations of motion for Lagrangian systems defined on  higher-order tangent bundles of a Lie group and finally, we will extend the results to the cases of variationally constrained problems.

\subsection{Unconstrained problem}\label{subsection2}
In 1983, it was shown by R. Skinner and R. Rusk \cite{Skinner-Rusk} that the dynamics of an autonomous
classical mechanical system with lagrangian $L: TQ\rightarrow \R$ can be  represented by a
hamiltonian presymplectic system on  the Whitney sum $TQ\oplus T^*Q$ (also called the Pontryaguin bundle).
 The Skinner-Rusk formulation can be briefly summarized as follows. Denoting
the projections of $TQ\oplus T^*Q\to TQ$ and $TQ\oplus T^*Q\to T^*Q$ by $pr_1$ and $pr_2$, respectively, we define
the presymplectic 2-form $\Omega=pr_2^*\omega_Q$ (where $\omega_Q$ is the canonical symplectic 2-form)
and the hamiltonian function $H: TQ\oplus T^*Q\to \R$ by
\[
H(v_q, \alpha_q)=\langle \alpha_q, v_q\rangle-L(v_q), \qquad \alpha_q\in T^*_qQ \hbox{   and   } v_q\in T_qQ\; .
\]
Then, one can consider the following equation
\[
i_X\Omega=dH\; ,
\]
which describes completely the equations of motion of the Lagrangian system. Moreover, if the Lagrangian is regular
then there exists a  unique solution $Z$  which is tangent to the graph of the Legendre map.
This formalism is quite interesting since it allows us to analyze the case of singular lagrangians in the same framework and, more important,  it also provides an appropriate
setting for a geometric approach to constrained variational optimization problems.

\subsubsection{The equations of motion}

Now, we will give an adaptation of the Skinner-Rusk algorithm to the case of higher-order theories on Lie groups.
We use the identifications
\begin{eqnarray*}
T^{(k)}G&\equiv&G\times k{\mathfrak g}\; ,\\
T^*T^{(k-1)}G&\equiv&G\times (k-1){\mathfrak g}\times k{\mathfrak g}^*\; .
\end{eqnarray*}
Consider the \emph{higher-order Pontryaguin bundle}
\[
W_0=T^{(k)}G\times_{T^{(k-1)}G}T^*T^{(k-1)}G\equiv G\times k{\mathfrak g}\times k{\mathfrak g}^*\; ,
\]
with induced projections
\begin{eqnarray*}
pr_1(g, {\bm \xi}, \xi^{(k-1)}, {\bm \alpha})&=&(g,  {\bm \xi}, \xi^{(k-1)})
\\
pr_2(g, {\bm \xi}, \xi^{(k-1)}, {\bm \alpha})&=&(g,  {\bm \xi}, {\bm \alpha})
\end{eqnarray*}
where, as usual,
${\bm \xi}=(\xi^{(0)}, \ldots, \xi^{(k-2)})\in (k-1){\mathfrak g}$ and ${\bm \alpha}=(\alpha_0,\ldots,\alpha_{k-1})\in k{\mathfrak g}^*$.

\begin{figure}[h]
$$\xymatrix{
  &&W_0=G\times k\mathfrak{g}\times k\mathfrak{g}^{*}\ar[lld]_{pr_1}  \ar[rrd]^{pr_2}&&\\
  G\times k\mathfrak{g} \ar[rrd]^{\tau^{(k-1,k)}_{G}} && && G\times (k-1)\mathfrak{g}\times k\mathfrak{g}^{*} \ar[lld]^{\pi_{G\times(k-1)\mathfrak{g} }} \\
  && G\times(k-1)\mathfrak{g} &&
   }$$
   %\caption{Skinner and Rusk formalism on Lie groups}
\end{figure}

For developing the Skinner and Rusk formalism it is only necessary to construct the  presymplectic 2-form $\Omega_{W_0}$ by
$\Omega_{W_0}=pr_2^*\omega_{G\times (k-1){\mathfrak g}}$ and the hamiltonian function  $H: W_0\to \R$ by
\[
H(g, {\bm \xi}, \xi^{(k-1)}, {\bm \alpha})=\sum_{i=0}^{k-1}\langle \alpha_i, \xi^{(i)}\rangle-L(g,  {\bm \xi}, \xi^{(k-1)})\; .
\]
Therefore
\begin{eqnarray*}
&&(\Omega_{W_0})_{(g, {\bm \xi}, \xi^{(k-1)}, {\bm \alpha})}\left( ({\bm \xi}_1,\xi_1^{(k)}, {\bm \nu}^1), ({\bm \xi}_2,\xi_2^{(k)}, {\bm \nu}^2)\right)=-\langle {\bm \nu}^1, {\bm \xi}_2\rangle + \langle {\bm \nu}^2, {\bm \xi}_1\rangle\\
&&\qquad+ \langle\alpha_0, [\xi^{(0)}_1, \xi^{(0)}_2]\rangle\;
=-\sum_{i=0}^{k-1}\left[\langle {\nu}^1_{(i)}, {\xi}^{(i)}_2\rangle - \langle {\nu}^2_{(i)}, {\xi}^{(i)}_1\rangle\right]+ \langle\alpha_0, [\xi^{(0)}_1, \xi^{(0)}_2]\rangle\; ,\label{omega-2}
\end{eqnarray*}
where ${\bm \xi}_a\in k{\mathfrak g}$, ${\bm \nu}^a\in {k\mathfrak g}^*$, and $\xi_a^{(k)}\in {\mathfrak g}$, $a=1, 2$.
Observe that $\xi_a^{(k)}$ does not appear on the right-hand side of the previous expression, as a consequence of the presymplectic character of $\Omega_{W_0}$.
Moreover, \begin{eqnarray*}
dH_{(g, {\bm \xi}, \xi^{(k-1)}, {\bm \alpha})}({\bm \xi}_2,\xi_2^{(k)}, {\bm \nu}^2)&=&
\langle -\pounds_g^*\left(\frac{\delta L}{\delta g}{(g, {\bm \xi}, \xi^{(k-1)})}\right), {\xi}^{(0)}_2\rangle\\
&&+\sum_{i=0}^{k-2}\langle \alpha_i-\frac{\delta L}{\delta \xi^{(i)}}{(g,{\bm \xi},  \xi^{(k-1)})}, {\xi}^{(i+1)}_2\rangle\\
&&+
 \langle {\bm \nu}^2, {\bm \xi}\rangle\; . \label{hami-3}
\end{eqnarray*}
Therefore, the \emph{intrinsic equations of motion} of a higher-order problem on Lie groups are now
\begin{equation}\label{eq-pr}
i_X\Omega_{W_0}=dH\; .
\end{equation}
If we look for a solution $X(g, {\bm \xi}, \xi^{(k-1)}, {\bm \alpha})=({\bm \xi}_1,\xi_1^{(k-1)}, {\bm \nu}^1)$ of Equation (\ref{eq-pr}) we deduce:
\begin{eqnarray*}
\xi_1^{(i)}&=&\xi^{(i)}, \quad 0\leq i\leq k-1\; ,\\
{\nu}^1_{(0)}&=&\pounds_g^*\left(\frac{\delta L}{\delta g}{(g, {\bm \xi}, \xi^{(k-1)})}\right)+ad_{\xi^{(0)}_1}\alpha_0\; ,\\
{\nu}^1_{(i+1)}&=&\frac{\delta L}{\delta \xi^{(i)}}{(g, {\bm \xi},  \xi^{(k-1)})}-\alpha_i, \quad 0\leq i\leq k-2\; ,
\end{eqnarray*}
and the constraint functions
\[
\alpha_{k-1}-\frac{\delta L}{\delta \xi^{(k-1)}}{(g,{\bm \xi},  \xi^{(k-1)})}=0\; .
\]
Observe that the coefficients $\xi^{k}_1$ are still undetermined.

An  integral curve of $X$, that is a curve of the type
\[
t\longrightarrow (g(t), \xi(t), \ldots, \xi^{(k-1)}(t), \alpha_0(t), \ldots, \alpha_{k-1}(t))\; ,
\]
must satisfy the following system of differential-algebraic equations (DAEs):
\begin{eqnarray}
\dot{g}&=&g\xi\; ,\\
\frac{d\xi^{(i-1)}}{dt}&=&\xi^{(i)}, \quad 1\leq i\leq k-1\; ,\\
\frac{d{\alpha}_0}{dt}&=&\pounds_g^*\left(\frac{\delta L}{\delta g}{(g, {\bm \xi}, \xi^{(k-1)})}\right)+ad^*_{\xi}\alpha_0\; ,\label{eq-11}\\
\frac{d{\alpha}_{i+1}}{dt}&=&\frac{\delta L}{\delta \xi^{(i)}}{(g,{\bm \xi},  \xi^{(k-1)})}-\alpha_i, \quad 0\leq i\leq k-2\; ,\label{eq-2}\\
\alpha_{k-1}&=&\frac{\delta L}{\delta \xi^{(k-1)}}{(g,{\bm \xi}, \xi^{(k-1)})}\; . \label{eq-l}
\end{eqnarray}
If $k\geq 2$,  combining Equation (\ref{eq-l}) with the (\ref{eq-2}) for $i=k-2$, we obtain
\[
\frac{d}{dt}\frac{\delta L}{\delta \xi^{(k-1)}}=\frac{\delta L}{\delta \xi^{(k-2)}}-\alpha_{k-2}\; .
\]
Proceeding successively, now with $i=k-3$  and ending with $i=0$ we obtain the following relation:
\[
 \alpha_0=\sum_{i=0}^{k-1}(-1)^i\frac{d^i}{dt^i}\frac{\delta L}{\delta \xi^{(i)}}\, .
 \]
 This last expression is also valid for $k\geq 1$.
 Substituting in the Equation (\ref{eq-11}) we finally deduce the \emph{ $k$-order trivialized Euler-Lagrange
equations}:
\begin{equation}\label{main}
(\frac{d}{dt}-ad^*_{\xi})\sum_{i=0}^{k-1}(-1)^i\frac{d^i}{dt^i}\frac{\delta L}{\delta \xi^{(i)}}=\pounds_g^*\left(\frac{\delta L}{\delta g}\right)\; .
\end{equation}

Of course if the Lagrangian $L: T^{(k)}G\equiv G\times k{\mathfrak g}\longrightarrow \R$ is left-invariant, that is
\[
L(g, \xi, \dot{\xi}, \ldots, \xi^{(k-1)})=L(h,\xi, \dot{\xi}, \ldots, \xi^{(k-1)})\; ,
\]
for all $g, h\in G$, then defining the reduced lagrangian
$
l: k{\mathfrak g}\longrightarrow \R$ by \[
l(\xi, \dot{\xi}, \ldots, \xi^{(k-1)})=L(e, \xi, \dot{\xi}, \ldots, \xi^{(k-1)})\; ,
\] we write Equations (\ref{main}) as
\begin{equation}\label{main-1}
(\frac{d}{dt}-ad^*_{\xi})\sum_{i=0}^{k-1}(-1)^i\frac{d^i}{dt^i}\frac{\delta l}{\delta \xi^{(i)}}=0\; ,
\end{equation}
 which are the \emph{$k$-order Euler-Poincar\'e equations} (see, for instance, \cite{GHMRV10}).

\subsubsection{The constraint algorithm}
Since $\Omega_{W_0}$ is presymplectic then (\ref{eq-pr}) has not solution along $W_0$ then it is necessary to identify the unique maximal submanifold $W_f$ along which (\ref{eq-pr}) possesses tangent solutions on $W_f$.
This final constraint submanifold $W_f$ is detected using the  Gotay-Nester-Hinds algorithm \cite{GoNesHinds78}.
This algorithm prescribes that $W_f$ is the limit of a string of sequentially constructed constraint submanifolds
\[
\cdots\hookrightarrow W_k\hookrightarrow \cdots \hookrightarrow
W_2\hookrightarrow W_1\hookrightarrow W_0\; .
\]
 where
\begin{eqnarray*}
W_i&=&\left\{x\in G\times k{\mathfrak g}\times k{\mathfrak g}^*\;\;
\big| \; \; dH (x)({\bm \xi}_1, \xi_1^{(k)}, {\bm \nu}^1)=0\;\right.\\
 &&\left. \ \forall ({\bm \xi}_1, \xi_1^{(k)}, {\bm \nu}^1)\in \left(T_x W_{i-1}\right)^{\perp} \;
\right\}
\end{eqnarray*}
with $i\geq 1$ and where
 \begin{eqnarray*}
 &&\left(T_x W_{i-1}\right)^{\perp}= \left\{ ({\bm \xi}_1, \xi_1^{(k)}, {\bm \nu}^1) \in (k+1){\mathfrak g}\times k{\mathfrak g}^* \; \big| \;
\Omega_{W_0}(x)(({\bm \xi}_1, \xi_1^{(k)}, {\bm \nu}^1),({\bm \xi}_2, \xi_2^{(k)}, {\bm \nu}^2)) = 0\right.
 \\&&\qquad\qquad\left. \; \ \forall \,  ({\bm \xi}_2, \xi_2^{(k)}, {\bm \nu}^2)\in T_x W_{i-1} \;
\right\}.
\end{eqnarray*}
where we are using the previously defined identifications.
If  this constraint algorithm stabilizes, i.e., there exists a positive integer $k\in \N$ such that $W_{k+1} = W_{k}$ and $\dim W_{k}\geq 1,$ then we will have at least a well defined solution $X$ on $W_{f} = W_{k}$ such that \[
\left(i_X\Omega_{W_0}=dH\right)_{| W_f}\; .
\]

From these definitions, we deduce that the first  constraint submanifold $W_1$ is defined by the vanishing of the constraint functions
\[
\alpha_{k-1}-\frac{\delta L}{\delta \xi^{(k-1)}}=0\; .
\]

 Applying the constraint algorithm we deduce that the following condition,  if $k>2$:
 \[
 \frac{\delta L}{\delta \xi^{(k-2)}}-\alpha_{k-2}=\frac{\delta^2 L}{\delta \xi^{(k-1)}\delta \xi^{(k-1)}}\xi_1^{(k)}+\sum_{i=0}^{k-2}\frac{\delta^2 L}{\delta \xi^{(k-1)}\delta \xi^{(i)}}\xi^{i+1}+
\pounds_g^*\left(\frac{\delta^2 L}{\delta \xi^{(k-1)}\delta g}\right)\xi\; .
\]
In the particular case $k=1$,
we deduce the equation
\[
 \pounds_g^*\left(\frac{\delta L}{\delta g}\right)+ad^*_{\xi}\alpha_0=\frac{\delta^2 L}{\delta \xi^2}\xi_1^{(1)}+
\pounds_g^*\left(\frac{\delta^2 L}{\delta \xi\delta g}\right)\xi\; .
\]
In both cases,  these equations impose restrictions over the remainder coefficients  $\xi_1^{(k)}$ of the vector field $X$.

If the bilinear  form $\frac{\delta^2 L}{\delta \xi^{(k-1)}\delta \xi^{(k-1)}}: {\mathfrak g}\times {\mathfrak g}\to \R$
defined by
\[
\frac{\delta^2 L}{\delta \xi^{(k-1)}\delta \xi^{(k-1)}}(g, {\bm \xi}, \xi^{(k-1)})(\xi, \tilde{\xi})=\frac{d}{dt}\Big|_{t=0}\frac{d}{ds}\Big|_{s=0}L(g, {\bm \xi}, \xi^{(k-1)}+t\xi+s\tilde{\xi})
\]
is nondegenerate, we have a special case when the constraint algorithm finishes at the first step $W_1$. More precisely, if we denote by $\Omega_{W_1}$ the restriction of the presymplectic 2-form $\Omega$ to $W_1$, then we have  the following result:

\begin{theorem}\label{theorem-1}
  $(W_1, \Omega_{W_1})$ is a symplectic manifold if and only if \begin{equation}
\frac{\delta^2 L}{\delta \xi^{(k-1)}\delta \xi^{(k-1)}}
\end{equation}
is nondegenerate.
\end{theorem}

\subsection{Constrained problem}\label{subsection3}

\subsubsection{The equations of motion}
The geometrical interpretation of constrained problems determined by a submanifold ${\mathcal M}$ of $G\times k{\mathfrak g}$, with inclusion $i_{\mathcal M}: {\mathcal M}\hookrightarrow G\times k{\mathfrak g}$ and a lagrangian function defined on it, $L_{\mathcal M}: {\mathcal M}\rightarrow \R$, is an  extension of the previous framework.
First, it is necessary to note that for constrained system, in this paper,  we understand a variational problem subject to constraints (vakonomic mechanics),  being this analysis completely different in the case of nonholonomic constraints (see \cite{Blo,CrtLeonMrtMrtz02,CrMe}).

Given the pair $( {\mathcal M}, L_{\mathcal M})$ we can define the space
\[
\overline{W}_0={\mathcal M}\times k{\mathfrak g}^*\, .
\]
Take the inclusion $i_{\overline{W}_0}: \overline{W}_0\hookrightarrow G\times k{\mathfrak g}\times k{\mathfrak g}^*$, then we can construct the following presymplectic form
\[
\Omega_{\overline{W}_0}=(pr_2\circ i_{\overline{W}_0})^*\Omega_{G\times (k-1){\mathfrak g}\times k{\mathfrak g}^*}\; ,
\]
and the function $\bar{H}:\overline{W}_0\to \R$ defined by
\[
\bar{H}(g, {\bm \xi}, \xi^{(k-1)}, {\bm \alpha})=\sum_{i=0}^{k-1}\langle \alpha_i, \xi^{(i)}\rangle-L_{\mathcal M}(g, {\bm \xi}, \xi^{(k-1)})\; ,
\]
where $(g, {\bm \xi}, \xi^{(k-1)})\in {\mathcal M}$.

With these two elements it is possible to write the following  presymplectic system:
\begin{equation}\label{poi}
i_X\Omega_{\overline{W}_0}=d\bar{H}\; .
\end{equation}
This then justifies the use of the following terminology.
\begin{definition}
The presymplectic Hamiltonian system $(\overline{W}_0,\Omega_{\overline{W}_0}, \bar{H})$ will be called the variationally constrained Hamiltonian system.
\end{definition}

 To characterize the equations we will adopt an ``extrinsic point of view", that is, we will work on the full space $W_0$ instead of in the restricted space $\overline{W_0}$ (see next section for an alternative  approach). Consider an arbitrary extension  $L: G\times k{\mathfrak g}\to \R$ of $L_{\mathcal M}: {\mathcal M}\to \R$. The main idea is to take into account that Equation (\ref{poi}) is equivalent to
\[
\left\{
\begin{array}{rcl}
i_X\Omega_{{W}_0}-d {H}&\in& \hbox{ann }T\overline{W}_0\; ,\\
X&\in&T\overline{W}_0\; ,
\end{array}
\right.
\]
where $\hbox{ann}$ denotes the annihilator of a distribution and $H$ is the function defined on Section \ref{subsection2}.

Assuming that ${\mathcal M}$ is determined by the vanishing of $m$-independent constraints
\[
\Phi^{A}(g, {\bm \xi}, \xi^{(k-1)})=0, \ 1\leq A\leq m\; ,
\]
then, locally,$
\hbox{ann }T\overline{W}_0=\hbox{span }\{ d\Phi^{A}\}\, ,
$
and therefore the previous equations is rewritten as
 \[
\left\{
\begin{array}{rcl}
i_X\Omega_{{W}_0}-d {H}&=& \lambda_{A}d\Phi^{A}\, ,\\
X(\Phi^{A})&=& 0\; ,
\end{array}
\right.
\]
where $\lambda_A$ are Lagrange multipliers to be determined.

If  $X(g, {\bm \xi}, \xi^{(k-1)}, {\bm \alpha})=({\bm \xi}_1,\xi_1^{(k)}, {\bm \nu}^1)$ then, as in the previous subsection, we obtain the following prescription about these coefficients:
\begin{eqnarray*}
\xi_1^{(i)}&=&\xi^{(i)}, \quad 0\leq i\leq k-1\, ,\\
{\nu}^1_{(0)}&=&\pounds_g^*\left(\frac{\delta L}{\delta g}-\lambda_{A}\frac{\delta \Phi^{A}}{\delta g}\right)+ad_{\xi^{(0)}_1}\alpha_0\; ,\\
{\nu}^1_{(i+1)}&=&\frac{\delta L}{\delta \xi^{(i)}}-\lambda_{A}\frac{\delta \Phi^{A}}{\delta \xi^{(i)}}-\alpha_i, \quad 0\leq i\leq k-2\; ,\\
0&=&\pounds_g^*\left(\frac{\delta \Phi^{A}}{\delta g}\right)\xi+\sum_{i=1}^{k-2}\frac{\delta \Phi^{A}}{\delta \xi^{(i)}}\xi^{(i+1)}+\frac{\delta \Phi^{A}}{\delta \xi^{(k-1)}}\xi_1^{(k )}\; ,\quad 1\leq A\leq m\, ,
\end{eqnarray*}
and the algebraic equations:
\begin{eqnarray*}
\alpha_{k-1}-\frac{\delta L}{\delta \xi^{(k-1)}}+\lambda_{A}\frac{\delta \Phi^{A}}{\delta \xi^{(k-1)}}&=&0\; , \\
\Phi^{A}&=&0\; .
\end{eqnarray*}

The integral curves of  $X$ satisfy the system of differential-algebraic equations with additional unknowns $(\lambda_{A})$:
\begin{eqnarray*}
\dot{g}&=&g\xi\; ,\\
\frac{d\xi^{(i-1)}}{dt}&=&\xi^{(i)}, \quad 1\leq i\leq k-1\; ,\\
\frac{d{\alpha}_0}{dt}&=&\pounds_g^*\left(\frac{\delta L}{\delta g}-\lambda_{A}\frac{\delta \Phi^{A}}{\delta g}\right)+ad^*_{\xi}\alpha_0\; ,\label{eq-1}\\
\frac{d{\alpha}_{i+1}}{dt}&=&\frac{\delta L}{\delta \xi^{(i)}}-\lambda_{A}\frac{\delta \Phi^{A}}{\delta \xi^{(i)}}-\alpha_i\; , \label{eq-2-1}\\
0&=&\pounds_g^*\left(\frac{\delta \Phi^{A}}{\delta g}\right)\xi+\sum_{i=1}^{k-2}\frac{\delta \Phi^{A}}{\delta \xi^{(i)}}\xi^{(i+1)}+\frac{\delta \Phi^{A}}{\delta \xi^{(k-1)}}\xi_1^{(k-1)}\\
\alpha_{k-1}&=&\frac{\delta L}{\delta \xi^{(k-1)}}-\lambda_{A}\frac{\delta \Phi^{A}}{\delta \xi^{(k-1)}}\; , \label{eq-l-2}\\
\Phi^{A}&=&0\; .
\end{eqnarray*}

As a consequence   we finally obtain the \emph{$k$-order trivialized constrained Euler-Lagrange
equations},
\begin{equation}\label{main-1-2}
(\frac{d}{dt}-ad^*_{\xi})\sum_{i=0}^{k-1}(-1)^i\frac{d^i}{dt^i}\left[\frac{\delta L}{\delta \xi^{(i)}}-\lambda_{A}\frac{\delta \Phi^{A}}{\delta \xi^{(i)}}\right]=\pounds_g^*\left(\frac{\delta L}{\delta g}-\lambda_{A}\frac{\delta \Phi^{A}}{\delta g}\right)\; .
\end{equation}

If the Lagrangian $L: T^{(k)}G\equiv G\times k{\mathfrak g}\longrightarrow \R$ and the constraints $\Phi^{A}: G\times k{\mathfrak g}\longrightarrow \R$, $1\leq A\leq m$ are left-invariant then defining the reduced lagrangian $l: k{\mathfrak g}\longrightarrow \R$ and the reduced constraints $\phi^{A}: k{\mathfrak g}\to \R$  we write Equations (\ref{main-1-2}) as
\begin{equation}\label{maiin-1-2}
(\frac{d}{dt}-ad^*_{\xi})\sum_{i=0}^{k-1}(-1)^i\frac{d^i}{dt^i}\left[\frac{\delta l}{\delta \xi^{(i)}}-\lambda_{A}\frac{\delta \phi^{A}}{\delta \xi^{(i)}}\right]=0\; .
\end{equation}

\subsubsection{The constraint algorithm}
As in the previous subsection it is possible to apply the Gotay-Nester algorithm to obtain a final constraint submanifold where we have at least a solution which is dynamically compatible.
The algorithm is exactly the same but applied to the equation (\ref{poi}).

Observe that the first constraint submanifold $\overline{W}_1$  is determined by the conditions
\begin{eqnarray*}
\alpha_{k-1}&=&\frac{\delta L}{\delta \xi^{(k-1)}}-\lambda_{A}\frac{\delta \Phi^{A}}{\delta \xi^{(k-1)}}\; , \\
\Phi^{A}&=&0\; .
\end{eqnarray*}
If we denote by $\Omega_{\overline{W}_1}$ the pullback of the presymplectic 2-form $\Omega_{\overline{W}_0}$ to $\overline{W}_1$, it is easy to prove the following

\begin{theorem}\label{theorem-2}
  $(\overline{W}_1, \Omega_{\overline{W}_1})$ is a symplectic manifold if and only if \begin{equation}
\left(
\begin{array}{ll}
\frac{\delta^2 L}{\delta \xi^{(k-1)}\delta \xi^{(k-1)}}&\frac{\delta \Phi^{A}}{\delta \xi^{(k-1)}}\\
\frac{\delta \Phi^{A}}{\delta \xi^{(k-1)}}&{\bm 0}
\end{array}
\right)
\end{equation}
is nondegenerate, considered as a bilinear form on the vector space ${\mathfrak g}\times \R^m$.
\end{theorem}

\section{An application: underactuated control systems on Lie groups}\label{section4}

A Lagrangian control system is underactuated if the
number of the control inputs is less than the dimension of the
configuration space (see \cite{bullolewis} and references therein). We assume that the controlled equations are trivialized where $L: G\times {\mathfrak g}\to \R$
\[
\frac{d}{dt}\left( \frac{\delta L}{\delta \xi}\right)-{ad}^*_{\xi}\left( \frac{\delta L}{\delta \xi}\right)-\pounds_g^*\frac{\partial L}{\partial g}=u_a e^a
\]
where we are  assuming that $\{e^a\}$ are independent elements on ${\mathfrak g}^*$ and $(u_a)$ are the admissible controls. Complete it to a basis $\{e^a, e^{A}\}$ of the vector space  ${\mathfrak g}^*$. Take its dual basis
$\{e_i\}=\{e_a, e_{A}\}$ on ${\mathfrak g}$ with bracket relations:
\[
[e_i, e_j]={\mathcal C}_{ij}^k e_k
\]
The basis $\{e_i\}=\{e_a, e_{A}\}$ induces coordinates $(y^a, y^{A})=(y^i)$ on ${\mathfrak g}$, that is, if $e\in {\mathfrak g}$ then
$e=y^i e_i=y^a e_a+y^{A} e_{A}$. In ${\mathfrak g}^*$, we have  induces coordinates $(p_a, p_{\alpha})$ for the previous fixed basis $\{e^i\}$

In these coordinates,  the equations of motion are rewritten as
\begin{eqnarray*}
\frac{d}{dt}\left(\frac{\partial L}{\partial y^a}\right)-{\mathcal C}_{ia}^jy^i\frac{\partial L}{\partial y^j}-\langle \pounds_g^*\frac{\delta L}{\delta g}, e_a\rangle &=&u_a\; ,\\
\frac{d}{dt}\left(\frac{\partial L}{\partial y^{A}}\right)-{\mathcal C}_{iA}^jy^i\frac{\partial L}{\partial y^j}-\langle \pounds_g^*\frac{\delta L}{\delta g}, e_{A}\rangle &=&0\; .
\end{eqnarray*}

With these equations we can study the optimal control problem that consists on finding  trajectories $(g(t), u^a(t))$ of state variables and control inputs satisfying the previous equations from given initial and final conditions $(g(t_0), y^i(t_0))$ and $(g(t_f), y^i(t_f))$, respectively, and
extremizing the functional
\[
{\mathcal J}=\int_{t_0}^{t_f} C(g(t), y^i(t),  u^a(t))\, dt
\]

Obviously (see \cite{Blo},\cite{CL91} and reference therein) the proposed optimal control problem is equivalent to a variational problem with second order constraints,
determined by the lagrangian $ \widetilde{L}:G\times 2{\mathfrak g}\to \R$ given, in the selected coordinates,  by
\[
\widetilde{L}(g, y^i,  \dot{y}^i)= C\left(g, y^i,
\frac{d}{dt}\left(\frac{\partial L}{\partial y^{a}}\right)-{\mathcal C}_{ia}^jy^i\frac{\partial L}{\partial y^j}-\langle \pounds_g^*\frac{\delta L}{\delta g}, e_{a}\rangle\right).
 \]
subjected to the second-order constraints
\[
\Phi^{A}(g,y^i, \dot{y}^i)=\frac{d}{dt}\left(\frac{\partial L}{\partial y^{A}}\right)-{\mathcal C}_{iA}^jy^i\frac{\partial L}{\partial y^j}-\langle \pounds_g^*\frac{\delta L}{\delta g}, e_{A}\rangle=0\; .
\]
 which determine the submanifold ${\mathcal M}$ of $G\times 2{\mathfrak g}$.

Observe that from the constraint equations we have that
\begin{eqnarray*}
\frac{\partial^2 L}{\partial y^{A}\partial y^{B}}\dot{y}^B+\frac{\partial^2 L}{\partial y^{A}\partial y^{b}}\dot{y}^b-{\mathcal C}_{iA}^jy^i\frac{\partial L}{\partial y^j}-\langle \pounds_g^*\frac{\delta L}{\delta g}, e_{A}\rangle &=&0\; .
\end{eqnarray*}

Therefore, assuming that the matrix $(W_{AB})=\left(\frac{\partial^2 L}{\partial y^{A}\partial y^{B}}\right)$ is regular we can write the constraint equations as
\begin{eqnarray*}
\dot{y}^B&=&-W^{BA}\left(\frac{\partial^2 L}{\partial y^{A}\partial y^{b}}\dot{y}^b-{\mathcal C}_{iA}^jy^i\frac{\partial L}{\partial y^j}-\langle \pounds_g^*\frac{\delta L}{\delta g}, e_{A}\rangle\right)\\
&=&G^{B}(g, y^i, \dot{y}^a)
\end{eqnarray*}
where $W^{BA}=(W_{BA})^{-1}.$
\begin{figure}[h]
$$\xymatrix{
  &&\overline{W}_0={\mathcal M}\times 2\mathfrak{g}^{*}\ar[lld]_{pr_1\circ i_{\overline{W}_0}}  \ar[rrd]^{pr_2\circ i_{\overline{W}_0}}&&\\
  {\mathcal M} \ar[rrd]^{(\tau^{(1,2)}_{G})|_{M}}  && && G\times\mathfrak{g}\times2\mathfrak{g}^{*} \ar[lld]^{\pi_{G\times {\mathfrak g}}} \\
  && G\times\mathfrak{g} &&
   }$$
   %\caption{Skinner and Rusk formalism}
\end{figure}

This means that we can identify $T{\mathcal M}\equiv G\times \hbox{span\; }\{(e_i, {\bm 0}, {\bm 0}), ({\bm 0}, e_i, {\bm 0}), ({\bm 0},{\bm 0}, e_a)\}$ where $(e_i, {\bm 0}, {\bm 0}), ({\bm 0}, e_i, {\bm 0}), ({\bm 0}, {\bm 0}, e_a)\in 3{\mathfrak g}$.

 Therefore, we can choose coordinates $(g, y^{i},\dot{y}^{a})$ on ${\mathcal M}$. This choice allows us to consider an ``intrinsic point view", that is, to work directly on $\overline{W}_0={\mathcal M}\times 2{\mathfrak g}^*$ avoiding the use of Lagrange multipliers.

 Define the restricted lagrangian $\widetilde{L}_{\mathcal M }$ by $\widetilde{L}_{\mathcal M } =
\widetilde{L}\mid_{\mathcal M}: {\mathcal M}\rightarrow\mathbb{R}$ and take induced  coordinates on $\overline
  {W}_0$ are  $\gamma=(g, y^{i}, \dot{y}^{a}, p_i, \tilde{p}_i)$. Consider the presymplectic 2-form on $\overline{W}_0$, $\Omega_{\overline{W}_0}=(pr_2\circ i_{\overline{W}_0})^*(\omega_{G\times {\mathfrak g}})$.

  Using the notation $(e_i)_{0}=(e_i, {\bm 0}, {\bm 0}; {\bm 0}, {\bm 0})\in 3{\mathfrak g}\times 2{\mathfrak g}^*$ and, in the same way $(e_i)_{1}=({\bm 0}, e_i, {\bm 0};  {\bm 0}, {\bm 0})$, $(e_a)_{2}=({\bm 0}, {\bm 0}, e_a; {\bm 0}, {\bm 0})$,  $(e^i)_{3}=( {\bm 0}, {\bm 0}, {\bm 0};   e^i, {\bm 0})$ and
  $(e^i)_{4}=( {\bm 0}, {\bm 0}, {\bm 0};  {\bm 0}, e^i)$ then the unique nonvanishing  elements on the expression of
  $\Omega_{\overline{W}_0}$ are:
  \begin{eqnarray*}
  (\Omega_{\overline{W}_0})_{\gamma}((e_i)_0, (e_j)_0)&=&{\mathcal C}_{ij}^kp_k\, ,\\
    (\Omega_{\overline{W}_0})_{\gamma}((e_i)_0, (e^j)_3)&=&-(\Omega_{\overline{W}_0})_{\gamma}((e^i)_3, (e_j)_0)=\delta^j_i\; ,\\
    (\Omega_{\overline{W}_0})_{\gamma}((e_i)_1, (e^j)_4)&=& -(\Omega_{\overline{W}_0})_{\gamma}((e^i)^4, (e_j)_1)=\delta_i^j\; .
   \end{eqnarray*}
Taking the dual basis   $(e^i)_{0}=(e^i, {\bm 0}, {\bm 0}; {\bm 0}, {\bm 0})\in 3{\mathfrak g}^*\times 2{\mathfrak g}$ and, in the same way $(e^i)_{1}=({\bm 0}, e^i, {\bm 0};  {\bm 0}, {\bm 0})$, $(e^a)_{2}=({\bm 0}, e^a,  {\bm 0}; {\bm 0}, {\bm 0})$,  $(e_i)_{3}=( {\bm 0}, {\bm 0}, {\bm 0};   e_i, {\bm 0})$ and
  $(e_i)_4=( {\bm 0}, {\bm 0}, {\bm 0};  {\bm 0}, e_i)$ we deduce that
  \[
  (\Omega_{\overline{W}_0})=(e^i)_0\wedge (e_i)_3+(e^i)_1\wedge (e_i)_4+\frac{1}{2}{\mathcal C}_{ij}^kp_k (e^i)_0\wedge (e^j)_0
  \]

Moreover
  \[
  \bar{H}=y^ip_i+ \dot{y}^a\tilde{p}_a+ G^A(g, y^i, \dot{y}^a)\tilde{p}_{A}-
\widetilde{L}_{\mathcal M }(g, y^i, \dot{y}^a).
\]
and, in consequence,
\begin{eqnarray*}
d\bar{H}&=&
-\langle \pounds_g^*\left(\frac{\delta \widetilde{L}_{\mathcal M }}{\delta g}+\tilde{p}_B\frac{\delta G^B}{\delta g}\right), e_i\rangle (e^i)_{0}+\left(p_i-\frac{\partial \widetilde{L}_{\mathcal M }}{\partial y^{i}}+\tilde{p}_B\frac{\partial G^B}{\partial y^{i}}\right)(e^i)_1\\
&&+\left(\tilde{p}_a-\frac{\partial \widetilde{L}_{\mathcal M }}{\partial \dot{y}^{a}}+\tilde{p}_B\frac{\partial G^B}{\partial \dot{y}^{a}}\right)(e^a)_2+y^i(e_i)_{3}+\dot{y}^a(e_a)_{4}+G^A(e_A)_{4}\; .
\end{eqnarray*}
The conditions for the integral curves $t\rightarrow (g(t), y^i(t), \dot{y}^a(t), p_{A}(t), \tilde{p}_{A}(t))$ of a vector field $X$ satisfying equations $i_X \Omega_{\overline{W}_0}=d\bar{H}$ are
 \begin{eqnarray}
  \frac{d g}{dt}&=&g (y^i(t)e_i)\\
  \frac{dy^a}{dt}&=&\dot{y}^a\\
 \frac{d y^A}{dt} &=& G^{A}(g, y^i, \dot{y}^a)\\
 \frac{d p_i}{dt} &=& \langle \pounds_g^*\left(\frac{\delta \widetilde{L}_{\mathcal M }}{\delta g}-\widetilde{p}_B\frac{\delta G^B}{\delta g}\right), e_i\rangle+{\mathcal C}_{ij}^kp_ky^j\\
 \frac{d \tilde{p}_{i}}{dt} &=& -p_i+\frac{\partial \widetilde{L}_{\mathcal M }}{\partial y^{i}}-\widetilde{p}_B\frac{\partial G^B}{\partial y^{i}}\\
\widetilde{p}_{a} &=& \frac{\partial \widetilde{L}_{\mathcal M }}{\partial \dot{y}^{a}}-\widetilde{p}_B\frac{\partial G^B}{\partial \dot{y}^{a}}=
\frac{\partial \widetilde{L}_{\mathcal M }}{\partial \dot{y}^{a}}+W^{BA}\widetilde{p}_B\frac{\partial^2 L}{\partial y^A\partial  {y}^{a}}\\
\end{eqnarray}
As a consequence we obtain the following set of differential equations:
\begin{eqnarray*}
\frac{d g}{dt}&=&g (y^i(t)e_i)\\
\frac{d y^A}{dt} &=& G^{A}(g, y^i, \dot{y}^a)\\
0&=&\frac{d^2}{dt^2}\left[\frac{\partial \widetilde{L}_{\mathcal M }}{\partial \dot{y}^{a}}-\widetilde{p}_B\frac{\partial G^B}{\partial \dot{y}^{a}}\right]-{\mathcal C}_{ia}^by^i\left(\frac{d}{dt}\left[\frac{\partial \widetilde{L}_{\mathcal M }}{\partial \dot{y}^{b}}-\widetilde{p}_B\frac{\partial G^B}{\partial \dot{y}^{b}}\right]\right)\\
&&-\frac{d}{dt}\left(\frac{\partial \widetilde{L}_{\mathcal M }}{\partial y^{a}}-\widetilde{p}_B\frac{\partial G^B}{\partial y^{a}}\right) +{\mathcal C}_{ia}^ky^i\left(\frac{\partial \widetilde{L}_{\mathcal M }}{\partial {y}^{k}}-\widetilde{p}_B\frac{\partial G^B}{\partial {y}^{k}}\right)\\
&&+\langle \pounds_g^*\left(\frac{\delta \widetilde{L}_{\mathcal M }}{\delta g}-\widetilde{p}_B\frac{\delta G^B}{\delta g}\right), e_a\rangle
-{\mathcal C}_{ia}^Cy^i\frac{d\tilde{p}_C}{dt}\\
0&=&\frac{d^2\widetilde{p}_A}{dt^2}+{\mathcal C}_{iA}^By^i\frac{d\widetilde{p}_B}{dt}
-{\mathcal C}_{iA}^ky^i\left[\frac{\partial \widetilde{L}_{\mathcal M }}{\partial {y}^{k}}-\widetilde{p}_B\frac{\partial G^B}{\partial {y}^{k}}\right]\\
&&-\frac{d}{dt}\left[\frac{\partial \widetilde{L}_{\mathcal M }}{\partial {y}^{A}}-\widetilde{p}_B\frac{\partial G^B}{\partial {y}^{A}}\right]
+\langle \pounds_g^*\left(\frac{\delta \widetilde{L}_{\mathcal M }}{\delta g}-\widetilde{p}_B\frac{\delta G^B}{\delta g}\right), e_A\rangle\\
&&
+{\mathcal C}_{iA}^by^i\left(\frac{d}{dt}\left[\frac{\partial \widetilde{L}_{\mathcal M }}{\partial \dot{y}^{b}}-\widetilde{p}_B\frac{\partial G^B}{\partial \dot{y}^{b}}\right]\right)
-{\mathcal C}_{iA}^by^i\left[\frac{\partial \widetilde{L}_{\mathcal M }}{\partial {y}^{b}}-\widetilde{p}_B\frac{\partial G^B}{\partial {y}^{b}}\right]
\end{eqnarray*}
which determine completely the dynamics.

If the matrix \[
\left(\frac{\partial^2 \widetilde{L}_{\mathcal M }}{\partial \dot{y}^{a}\partial\dot{y}^b}\right)
\]
is regular then we can write the previous equations as a explicit system of third-order differential equations. It is easy to show that this regularity assumption is equivalent to the condition that the constrain algorithm  stops at the first constraint submanifold  $\overline{W}_1$ (see \cite{Maria},\cite{ldm}, \cite{CoMa10} and reference therein for more details).

\subsection{Example: optimal control of an underactuated rigid body}

We consider the motion of a rigid body where the configuration space is the Lie group $G=SO(3)$ (see \cite{BHLS,Leok}). Therefore,  $TSO(3)\simeq SO(3)\times\mathfrak{so}(3),$ where $\mathfrak{so}(3)\equiv \R^3$ is the Lie algebra of the Lie group $SO(3).$ The Lagrangian function for this system is given by $L:SO(3)\times\mathfrak{so}(3)\rightarrow\R,$  $$L(R,\Omega_1,\Omega_2,\Omega_3)=\frac{1}{2}\left(I_1\Omega_1^2+I_2\Omega_2^2+I_3\Omega_3^2\right).$$

Now, denote by $t\to R(t)\in SO(3)$ a curve. The columns of the matrix $R(t)$ represent the directions of the principal axis of the body at time $t$ with respect to some reference system.
Now, we consider the following control problem.
First, we have the reconstruction equations:
$$\dot{R}(t)=R(t)
\left(
  \begin{array}{ccc}
    0& -\Omega_3(t) & \Omega_2(t) \\
    \Omega_3(t) & 0 & -\Omega_1(t) \\
    -\Omega_2(t) & \Omega_1(t) & 0 \\
  \end{array}
\right)=R(t)\left(\Omega_1(t)E_1+\Omega_2(t)E_2+\Omega_3(t)E_3\right)$$
where \[E_1:=\left(
               \begin{array}{ccc}
                 0 & 0 & 0 \\
                 0 & 0 & -1 \\
                 0 & 1 & 0 \\
               \end{array}
             \right), \qquad E_2:=\left(
                                   \begin{array}{ccc}
                                      0 & 0 & 1 \\
                                      0 & 0 & 0 \\
                                      -1 & 0 & 0 \\
                                    \end{array}
                                  \right), \qquad E_3:=\left(
                                                         \begin{array}{ccc}
                                                           0 & -1 & 0 \\
                                                           1 & 0 & 0 \\
                                                           0 & 0 & 0 \\
                                                         \end{array}
                                                       \right)\]
and the equations for the angular velocities $\Omega_i$ with $i=1,2,3$:
\begin{eqnarray*}
I_1\dot{\Omega}_1(t)&=&(I_2-I_3)\Omega_2(t)\Omega_3(t)+u_1(t)\\
I_2\dot{\Omega}_2(t)&=&(I_3-I_1)\Omega_3(t)\Omega_1(t)+u_2(t)\\
I_3\dot{\Omega}_3(t)&=&(I_1-I_2)\Omega_1(t)\Omega_2(t)
\end{eqnarray*}
where $I_1,I_2,I_3$ are the moments of inertia and $u_1, u_2$ denotes the applied torques playing the role of controls of the system.

The optimal control problem for the rigid body consists on finding  the trajectories $(R(t), \Omega(t), u(t))$  with fixed initial and final conditions
$(R(t_0), \Omega(t_0)),$ $(R(t_f), \Omega(t_f))$ respectively and minimizing the cost functional
$$\mathcal{A} =\int_0^T \mathcal{C}(\Omega, u_1, u_2) dt= \int_{0}^{T} \left[c_1(u_1^2+u_2^2)+c_2(\Omega_1^2+\Omega_2^2+\Omega_3^2)\right]\, dt,$$
with $c_1,c_2\geq 0.$
The constants $c_1$ and $c_2$ represent  \textbf{weights} on the cost functional. For instance, $c_1$ is the weight in the cost functional measuring  the fuel expended by an attitude manoeuver of a spacecraft modeled by the rigid body  and $c_2$ is the weight
 given  to  penalize high angular velocities.

This optimal control problem is equivalent to solve the following variational problem with constraints (\cite{Blo},\cite{CL91}),
$$\min \widetilde{\mathcal{J}} = \int_{0}^{T} \widetilde{L}(\Omega, \dot\Omega)dt$$  subject to constraints $I_3\dot\Omega_3-(I_1-I_2)\Omega_1\Omega_2=0,$
where
\[
\widetilde{L}(\Omega, \dot\Omega) = \mathcal{C}\left(\Omega, I_1\dot\Omega_1 - (I_2-I_3)\Omega_2\Omega_3 ,I_2\dot\Omega_2 - (I_3-I_1)\Omega_3\Omega_1 \right)\; .
\]
Thus, the submanifold $\mathcal{M}$ of $G\times 2\mathfrak{so}(3)$, is given by
$$\mathcal{M}=\{(R,\Omega,\dot{\Omega})\mid \dot{\Omega}_3=\left(\frac{I_1-I_2}{I_3}\right)\Omega_1\Omega_2\}.$$

We consider the submanifold $\overline{W}_0=\mathcal{M}\times 2\mathfrak{so}^{*}(3)$ with induced coordinates $$(g,\Omega_1,\Omega_2,\Omega_3,\dot{\Omega}_1,\dot{\Omega}_2, p_1, p_2, p_3, \tilde{p}_1, \tilde{p}_2, \tilde{p}_3).$$

Now, we consider the restriction $L_{\mathcal{M}}$ given by $$\tilde{L}_{\mathcal{M}}=c_1\left[\left(I_1\dot\Omega_1-(I_2-I_3)
\Omega_2\Omega_3\right)^2+\left(I_2\dot\Omega_2-(I_3-I_1)\Omega_3\Omega_1\right)^2\right]
+c_2\left(\Omega_1^2+\Omega_2^2+\Omega_3^2\right)\; .$$
For simplicity we denote by $G^{3}=\frac{I_1-I_2}{I_3}\Omega_1\Omega_2$.

\begin{center}
\begin{figure}[h]
  \includegraphics[width=12cm]{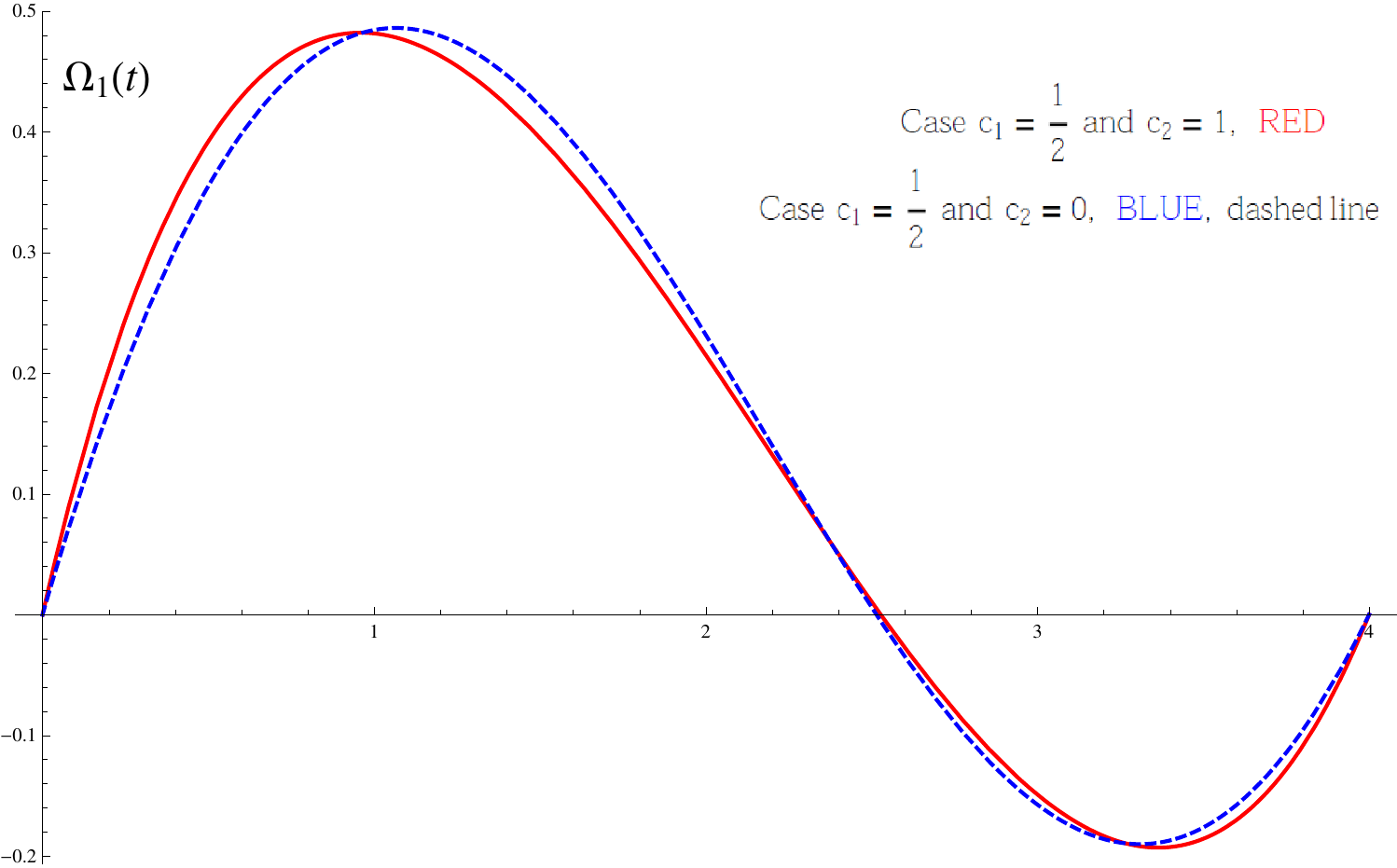}\\ $\,$\\\includegraphics[width=12cm]{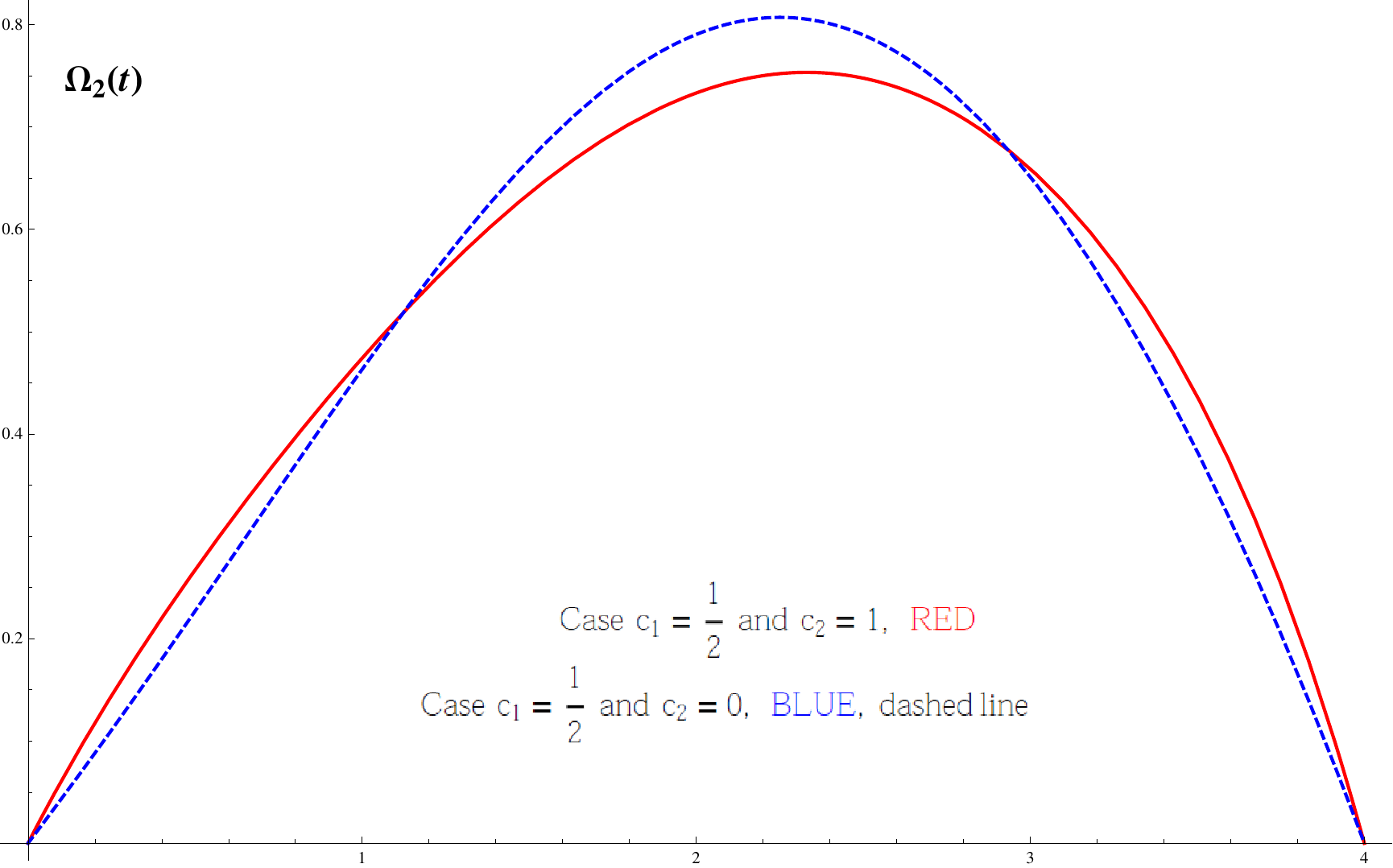}
  \caption{Angular velocity values for initial conditions satisfying $\Omega_i(0)=\Omega_i(4)=0$, $i=1, 2$ and fixed values of $R(0)$ and $R(4)$.}\label{aaa}
\end{figure}
\end{center}

\begin{center}
\begin{figure}[h]
  \includegraphics[width=6.5cm]{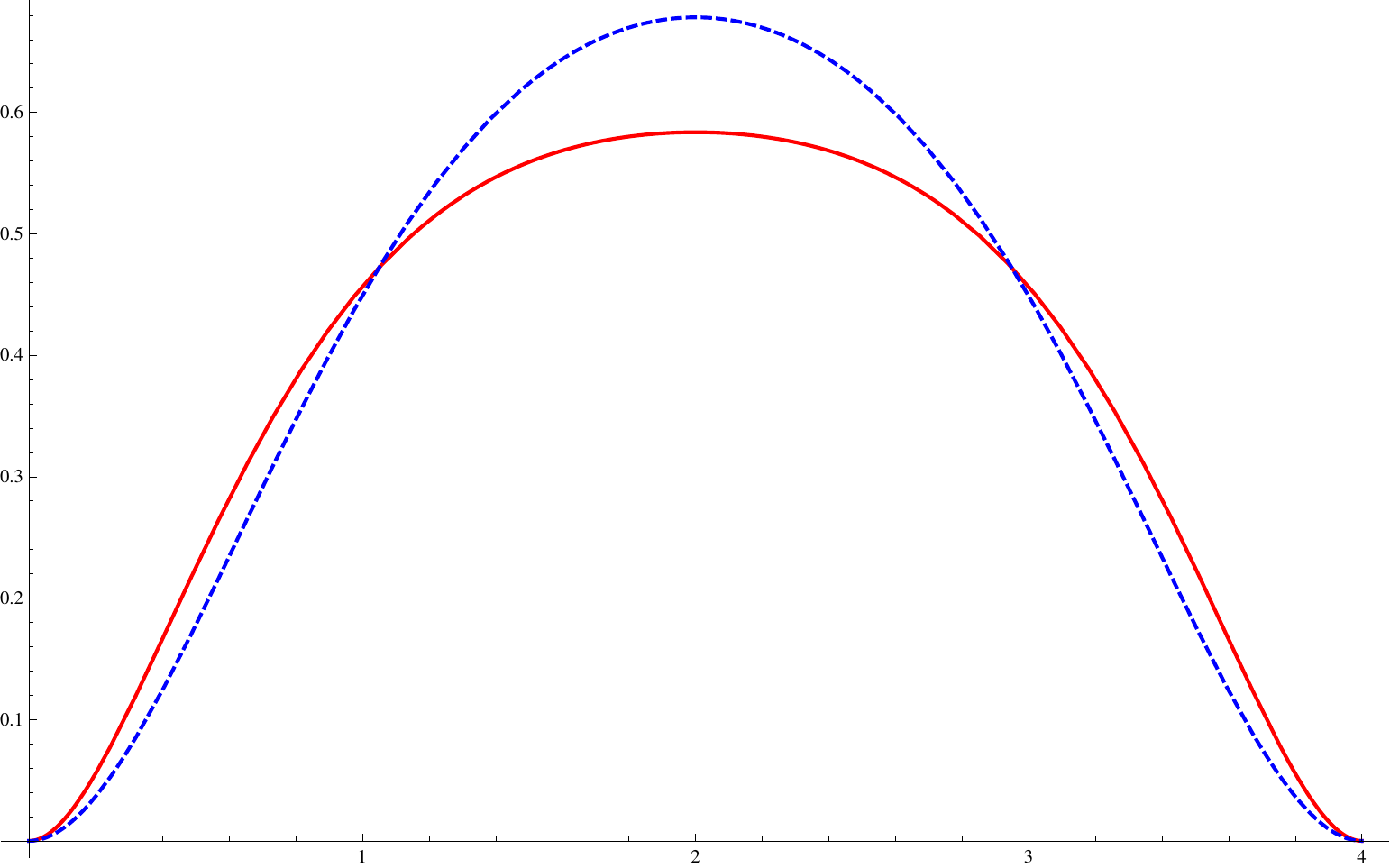}\ \includegraphics[width=6.5cm]{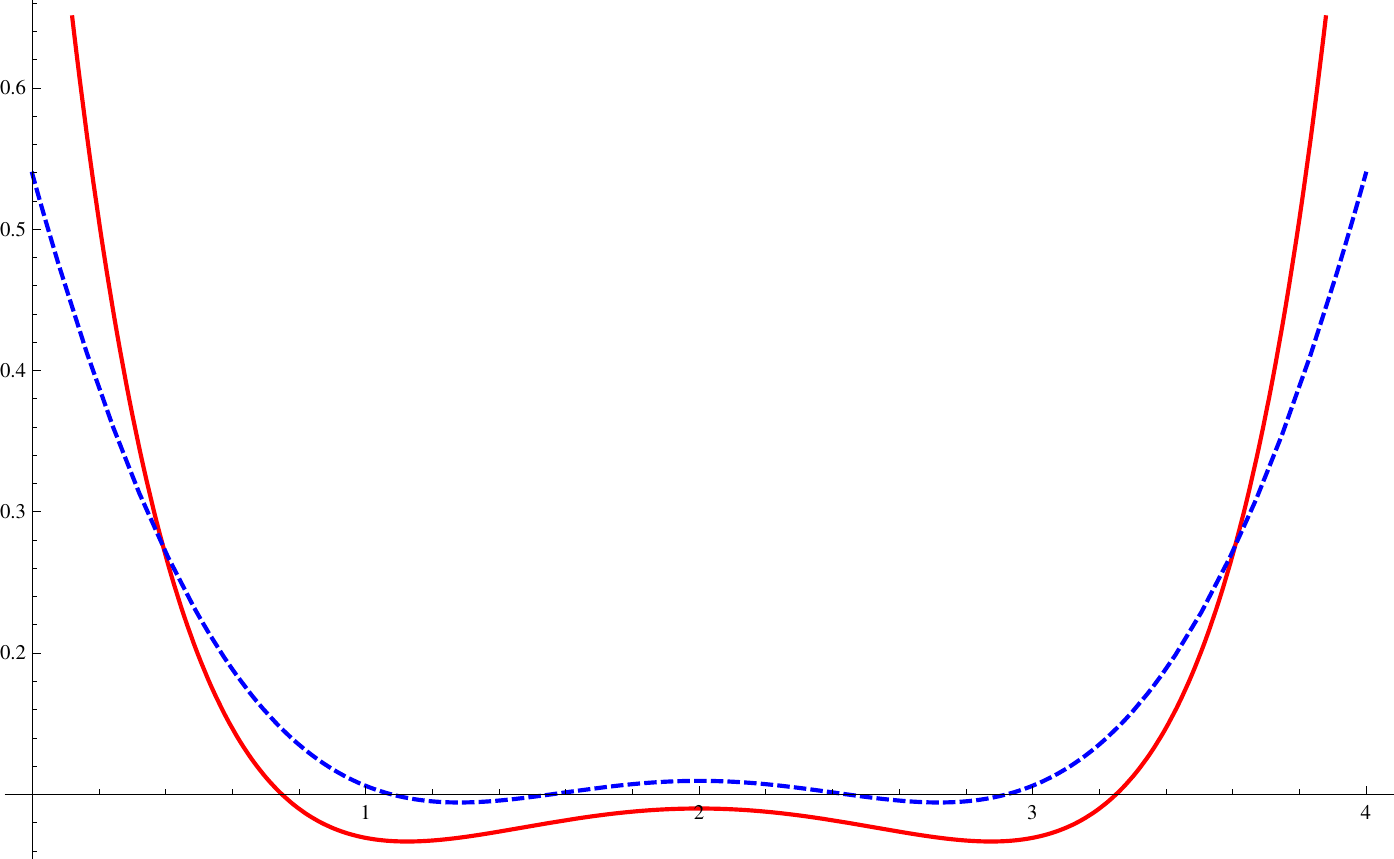}
  \caption{Comparison of the functions $1/2\left(\Omega^2_1(t)+\Omega^2_2(t)+\Omega^2_3(t)\right)$ (left) and $1/2\left(u^2_1(t)+u^2_2(t)\right)$ (right) in both cases}\label{aab}
\end{figure}
\end{center}

Then, we can write the equations of motion of the optimal control for this underactuated system. For simplicity, we consider the particular case $I_1=I_2=I_3=1$ then the equations of motion of the optimal control system are:
\begin{eqnarray*}
\Omega _2(t) \frac{d\tilde{p}_3}{dt}-2 \left( c_2 \frac{d\Omega_1}{dt}+c_1 \Omega_3(t) \frac{d^2\Omega_2}{dt^2}-c_1\frac{d^3 \Omega _1}{dt^3}\right)&=&0\\
-\Omega _1(t) \frac{d\tilde{p}_3}{dt}-2 \left(c_2 \frac{d\Omega_2}{dt}-c_1 \Omega_3(t) \frac{d^2\Omega _1}{dt^2}-c_1\frac{d^3 \Omega _2}{dt^3}\right)&=&0\\
\frac{d^2\tilde{p}_3}{dt^2}-2 c_2 \frac{d\Omega_3}{dt}-2c_1\Omega_2(t)\frac{d^2\Omega_1}{dt^2}
+2c_1\Omega_1(t)\frac{d^2\Omega_2}{dt^2}&=&0\\
\frac{d\Omega_3}{dt}&=&0
\end{eqnarray*}
If we consider the rigid body as a model of a spacecraft then we observe that this particular cost function is taking into account both, the fuel expenditure  ($c_1$) and also is trying to minimize the overall angular velocity ($c_2$).
In Figures (\ref{aaa}) and (\ref{aab}) we compare their behavior in two particular cases: $c_1=1/2$, $c_2=1/2$;  and $c_1=1/2$, $c_2=0$.

%The most typical case is of course the problem of minimize the total fuel expenditure, that is $c_1=1$ and $c_2=0$. (Case 2). Now,  the explicit system of differential equations is
%\begin{eqnarray*}
%\Omega _2(t) \frac{d\tilde{p}_3}{dt}-2\left(\frac{d^2\Omega _2}{dt^2}\Omega _3(t)-\frac{d^3 \Omega _1}{dt^3}\right)&==&0,\\
%-\Omega _1(t) \frac{d\tilde{p}_3}{dt}+2\left(\frac{d^2\Omega _1}{dt^2}\Omega _3(t)+\frac{d^3 \Omega _2}{dt^3}\right)&==&0,\\
%\frac{d^2\tilde{p}_3}{dt^2}&==&0,\\
%\frac{d\Omega_3}{dt}&==&0
%\end{eqnarray*}
In all cases we additionally have the reconstruction equation
$$\dot{R}(t)=R(t)\left(\Omega_1(t)E_1+\Omega_2(t)E_2+\Omega_3(t)E_3\right)\;. $$
with boundary conditions  $R(t_0)$ and $R(t_f)$.

The case   $c_1=0$ and $c_2=1$, that is, we only try to minimize the overall angular velocity (see \cite{spindler} for the fully-actuated case) is singular.   We obtain the following system
\begin{eqnarray*}
\Omega_2(t) \frac{d\tilde{p}_3}{dt}-2\frac{d\Omega_1}{dt}&=&0,\\
-\Omega _1(t) \frac{d\tilde{p}_3}{dt}-2\frac{d\Omega_2}{dt}&=&0,\\
\frac{d^2\tilde{p}_3}{dt^2}-2\frac{d\Omega_3}{dt}&=&0,\\
\frac{d\Omega_3}{dt}&=&0\; .
\end{eqnarray*}
Observe that in this case it is not possible to impose arbitrary  boundary conditions $(R(t_0), \Omega(t_0))$ and $(R(t_f), \Omega(t_f))$ although it is always possible to find a trajectory verifying initial and final attitude conditions $R(t_0)$ and $R(t_f)$.

\section{Conclusions and future work}

We have defined following an intrinsic point of view the equations of motion for variational higher-order lagrangian problems  with constraints. As a particular case, we obtain the higher-order Euler-Poincar\'e equations (see \cite{GHMRV10}). As an interesting application we deduce the equations of motion for optimal control of underactuated mechanical systems defined on Lie groups. These systems appear in numerous engineering and scientific fields, as for instance in  astrodynamics. In this sense we study the attitude control of a satellite modeled as a classic rigid body.

These techniques admits an easy generalization for the case of discrete systems.
As an illustration , consider  the second order tangent bundle of a Lie group   left trivialized as $T^{(2)}G\simeq G\times 2\mathfrak{g}$. We choose its natural discretization as three copies of the group (we recall that the prescribed discretization of a Lie algebra $\mathfrak{g}$ is its associated Lie group $G$). Consequently, we develop the discrete Euler-Poincar\'e equations for the discrete Lagrangians defined on  $L_d:G\times G\times G\rightarrow\mathbb{R}$. Define $W_k=g_k^{-1}g_{k+1}$. Taking variations for $W_k,$ where we denote $\Sigma_k=g_k^{-1}\delta g_k$, we obtain
\begin{eqnarray*}
\delta W_k&=&-g_k^{-1}\delta g_kg_{k}^{-1}g_{k+1} + g_k^{-1}\delta g_{k+1}\\
&=&-\Sigma_k W_k+ g_k^{-1}g_{k+1}g_{k+1}^{-1}\delta g_{k+1}\\
&=&-\Sigma_k W_k+ W_k\Sigma_{k+1},
\end{eqnarray*}
where $g_k,W_k\in G$ and $\Sigma_k\in\mathfrak{g}$.

The equations of motion are the critical paths of the discrete action $$\min\sum_{k=0}^{N-2}L_d(g_k,W_k,W_{k+1})$$ with boundary conditions $\Sigma_0=\Sigma_1=\Sigma_{N-1}=\Sigma_N=0$ since we are assuming that $g_0$, $g_1$, $g_{N-1}$ and $g_N$ fixed.  Therefore, after some computations we can obtain the equations,
%
%&&\delta\sum_{k=2}^{N-2}L_d(g_k,W_k,W_{k+1})=\sum_{k=2}^{N-2}L_d(g_k,W_k,W_{k+1})\delta g_k\\
%&&+D_3L_d(g_k,W_k,W_{k+1})(-\Sigma_{k+1}W_{k+1}+W_{k+1}\Sigma_{k+2})\\
%&&+D_2L_d(g_k,W_k,W_{k+1})\delta W_{k}=\\
\begin{eqnarray*}
&&l_{g_{k-1}}^{*}D_1L_d(g_{k-1},W_{k-1},W_k)+l_{W_{k-1}}^{*}D_2L_d(g_{k-1},W_{k-1},W_{k})\\
&&-r_{W_k}^{*}D_2L_d(g_k,W_k,W_{k+1})-r_{W_k}^{*}D_3L_d(g_{k-1},W_{k-1},W_k)\\
&&+l_{W_{k-1}}^{*}D_3L_d(g_{k-2},W_{k-2},W_{k-1})=0
\end{eqnarray*}
which are the discrete second-order Euler-Lagrange equations.

Moreover, in a future paper we will generalize the presented construction of  higher-order Euler-Lagrange equations to the case of Lie algebroids. This abstract approach will allows us to intrinsically derive the equations of motion for different cases as, for instance,  higher Euler-Poincar\'e equations,  Lagrange-Poincar\'e equations and the reduction by morphisms in a unified way. We will generalize the notion of higher-order tangent bundle to the case of higher-order Lie algebroids (or, more generally, anchored bundles) using equivalence classes of admissible curves and extending the ideas introduced in \cite{singlag}.

We will analyze in a future paper these and other related aspects.

\end{document}